\documentstyle[aps,prl,epsf]{revtex} 
\begin{document}
\draft
\tighten
\twocolumn[\hsize\textwidth\columnwidth\hsize\csname
@twocolumnfalse\endcsname
\title{Spin-based quantum computation in multielectron quantum dots}
\author{Xuedong Hu and S. Das Sarma}
\address{Department of Physics, University of Maryland, College
Park, MD 20742-4111}
\date{\today}
\maketitle
\begin{abstract}
In a quantum computer the hardware and software are intrinsically
connected because the quantum Hamiltonian (or more precisely its time
development) is the code that runs the computer.  We
demonstrate this subtle and crucial relationship by considering
the example of electron-spin-based solid state quantum computer in
semiconductor quantum dots.  We show that multielectron quantum dots
with one valence electron in the outermost shell do not behave simply
as an effective single spin system unless special conditions are
satisfied.  Our work compellingly demonstrates that a delicate synergy
between theory and experiment (between software and hardware) is
essential for constructing a quantum computer.  
\end{abstract}
\vskip2pc] 
\narrowtext

Ever since the pioneering work 
on quantum computation and quantum error correction
\cite{Reviews,Shor,Grover,error_correction}, 
there have been many 
proposed quantum computer (QC) hardware architectures based on 
different quantum systems \cite{special}, 
such as trapped ions \cite{iontrap}, 
cavity QED \cite{cavityQED}, 
liquid state NMR \cite{NMR}, 
nuclear spins in solids \cite{BKane1}, 
electron spins \cite{LD,Imam,Vrijen}, 
superconducting Josephson junctions \cite{Schon}, 
electrons on He surface \cite{Platzman},
etc.  Currently, experimental progress 
has mostly occurred in proposals based on atomic, optical,
and NMR physics.
Many solid state proposals have remained in the
model stage because of the immense experimental difficulties.
To help overcome these difficulties, 
more theoretical work is needed to explore the optimal 
operating regimes, figure out the operational constraints and
tolerances, and discover potential sources of errors, just 
to name a few directions \cite{DiVince,BLD,HD,Inhomo}.   
While the optical and atomic physics based architectures have been
crucial in demonstrating the proof of principle for quantum
computation, it is generally believed that solid state QC
architectures, with their obvious advantage of controllable scale-up
possibilities, offer the most promising potential for realistic large
scale QC hardwares.  The fundamental problem plaguing the solid state
QC architectures has been the fact that the basic quantum bit (qubit),
the QC building block, has not been compellingly demonstrated in any
solid state QC architectures, although there is no reason to doubt
that they exist in nature.  Thus, the construction of successful QC
hardwares has faced the somewhat embarrassing dichotomy: the
architectures (ion traps, etc.) demonstrating existence
of quantum bits cannot be easily scaled up, while the architectures
(solid state QCs) which may be easily scaled up have not yet
experimentally demonstrated quantum bits!

Quantum computation with fermionic spins is considered to be a
potentially promising prospect for solid state quantum computers
\cite{BKane1,LD,Imam,Vrijen,DiVincenzo}.  
Among the many proposed
solid state QC architectures the spin quantum computer
has several intrinsic advantages: (1) A fermionic spin, being a
quantum two-level system, is a natural qubit with its
spin up and down states; (2) it is fairly straightforward to carry
out single-qubit operations on spin up and down levels by applying
suitable magnetic fields (or through a purely 
exchange-based scheme \cite{DiVince}); 
(3) two-qubit operations
can, in principle, be carried out rather easily (in theory, at least)
by using the exchange interaction between two neighboring spins; (4)
quantum spin is fairly robust and does not decohere easily (typical
electron spin relaxation times in solids are many orders of magnitude
longer \cite{Jaro} 
than the momentum relaxation time)---in
particular, electron spin relaxation times could be microseconds
in semiconductors \cite{Awschalom}.

Our work presented in this paper deals with a crucial
aspect of solid state spin qubits which has so far been neglected in
the literature.  
The intrinsic advantages of spin based solid state quantum
computation have led to several concrete proposals for using electron
spins [in semiconductor quantum dots (QD) or in donor impurity atoms]
as qubits in semiconductor based solid state QC
architectures \cite{BKane1,LD,Imam,Vrijen}.  
One exciting proposal \cite{LD} 
deals with one electron spin
per quantum dot working as a qubit, with two coupled spins on two
neighboring dots (forming a QD molecule
\cite{Blick,Oosterkamp,Kotlyar}) providing two-qubit operations
through the inter-dot electronic exchange coupling.  The electron
spin on shallow donor states in semiconductors, while differing in
some details with the QD spin qubit architecture, still exploits the
idea of only one effective spin-1/2 fermion (i.e. one electron) per
donor state participating in the quantum computation \cite{Vrijen}.  
At first sight this idea of single electron in a dot
may seem far-fetched because an array of semiconductor QDs,
even under the most advanced growth and nanofabrication constraints, 
are likely to have more than a single electron on each dot
\cite{Ashoori}.   
However, the idea of one effective electron spin per quantum dot
working as a qubit is not as
crazy as it may seem at first sight.  In particular, QD
electronic states are, similar to real atomic electronic states,
naturally divided into quantized shells (i.e. S, P, D, F, etc.)
corresponding to the quantization of the orbital motion
\cite{Ashoori}.
Furthermore, the orbital excitation energy 
in a small QD is much higher than the spin flip energy for realistic
fields.  The single electron spin per quantum dot idea is therefore
based on the closed shell principle, where what is required is just
one ``valence'' electron per quantum dot in the outermost ``open''
shell.  The underlying idea here is that the closed shell electrons
(equivalent to the core electrons in atoms) are ``inert''
and could be ignored as far as qubit dynamics goes, and the
unoccupied states are energetically too unfavorable to be involved
as well.  This principle
in a different context has, in fact, worked for trapped-ion
quantum computation \cite{special} 
where ``valence'' type ionic
orbital states are manipulated as qubits, and the filled inner shell
states are inert and are ignored.
\vspace*{-0.1in}
\begin{figure}
\centerline{
\epsfxsize=3in
\epsfbox{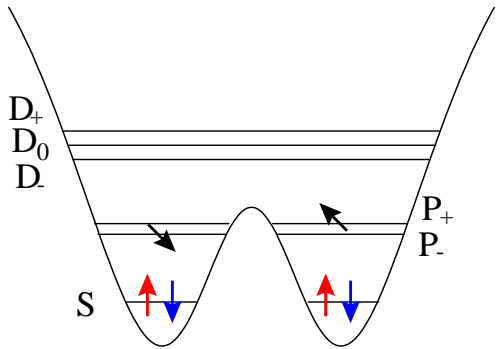}}
\vspace*{0.1in}
\protect\caption[6-electron schematic]
{\sloppy{
Here we show a schematic of six electrons in a double dot.  Four
of the electrons will occupy the four lowest spin orbitals (two S 
orbitals with two spin orientations), thus fill up the S shell
states.  The other two electrons can be in any one of the rest ten P
and D spin orbitals in our calculations.
}
}
\label{fig1}
\end{figure}

The all-important question for quantum dot electron spin quantum
computation is therefore the extent to which this same scenario 
applies, i.e., both the {\it inert} filled core of a quantum dot
and the outer-shell unoccupied orbital 
states can be ignored for quantum
computation because they do not affect the qubit
dynamics either for single-qubit or for exchange-mediated two-qubit
operations. 
The answer to this question is nontrivial and non-obvious
because the confinement potential in quantum dots is very different
from and much softer than that for real atoms.  In addition, the
gated circular QDs are essentially two dimensional and the Fock-Darwin
states (two dimensional electron eigenstates in a magnetic
field and a harmonic confinement)
are isotropic, unlike the three dimensional anisotropic
atomic states.  We address 
this crucial issue of electron-spin-based QD quantum
computation by accurately calculating the energy levels and exchange
couplings in multielectron QD molecules where two
semiconductor (GaAs) quantum dots, each with three electrons, are
used as the fundamental building block of the quantum computer
architecture (Fig.~\ref{fig1}).  We perform a configuration
interaction (CI) calculation with a Hartree-Fock basis.  
Specifically, we expand the single electron
states in a basis including all 12 S, P, and D Fock-Darwin states
located at the two potential minima.  This leads to $12 \times 2=24$
Hartree-Fock spin orbitals (each spatial orbital has 2 spin
orientation).  We include both singly and doubly excited 6-electron
states in the CI basis, and solve the Schr\"{o}dinger
equation by expanding on the 6-electron Slater basis
(Zeeman coupling has been neglected in this calculation): 
\begin{equation}
H(1,\ldots,6) \sum_i c_i \Psi_i(1,\ldots,6) = E \sum_i c_i 
\Psi_i(1,\ldots,6) \,,
\end{equation}
where $H$ is the 6-electron Hamiltonian including kinetic and
potential energy and electron Coulomb interaction.  As our theory is
based on a sophisticated quantum chemistry approach \cite{Cook}, our
results should have general qualitative and semi-quantitative
validity.  There have been several recent theoretical calculations
of the ground state spin polarization properties of multielectron
quantum dot systems using the density functional theory
\cite{Martin,Rossler,Yakimenko}.  For the purpose of quantum
computation of interest to us in this paper, however, the knowledge
of the excited states is crucial in determining whether a particular
number of electrons can serve as an effective qubit---in particular,
we need an accurate evaluation of the singlet-triplet energy
splitting in the system.  Such excited state information is beyond
the scope of density functional theories which are restricted to
ground states only.  The quantum-chemical CI calculations we present
in this paper are particularly well-suited in dealing with the low
lying excited states and in providing information about the exchange
splitting in the system (in contrast to ground state density
functional theories).

\begin{figure}
\centerline{
\epsfxsize=3.6in
\epsfbox{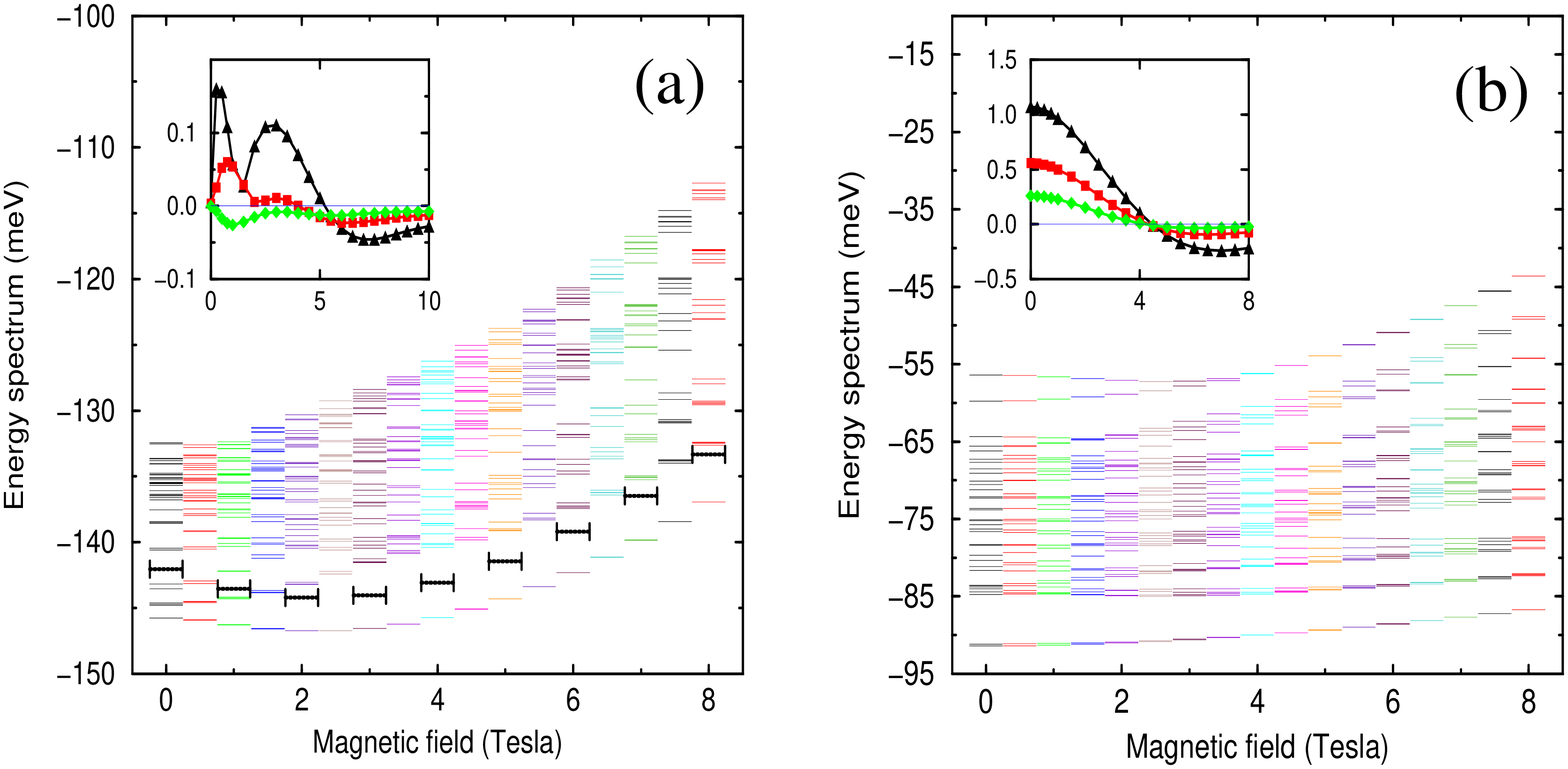}}
\vspace*{0.1in}
\protect\caption[Energy spectra of a 6-electron double dot]
{\sloppy{
In (a) we plot the energy spectra (lowest 40 states) of a 
particular 6-electron horizontal double dot as a function 
of an applied magnetic field along the z direction.  
The quantum dot widths (Gaussian confinement widths) are 30 nm 
in radius.  The distance between the two confinement potential 
minima is 40 nm. The central barrier height is 30 meV (with 
effective height of 19.28 meV).  For a more detailed description 
of the Gaussian confinement and barrier we use here, and a 
description of the horizontal quantum dots we study, see \cite{HD}.  
The thick dotted black lines with risers are
the ground state energies of our restricted Hartree-Fock 
self-consistent states, plotted here as a starting point to
compare our CI results with.  The inset shows how the splitting
of the lowest singlet and triplet states varies with the 
external magnetic field (at three different effective barrier
heights of 15.10, 19.28, and 23.65 meV).  
For the purpose of comparison,
in (b) we plot the energy spectra (lowest 36 states) of a 
particular two-electron horizontal double dot as a function 
of an applied magnetic field along the z direction.  
Here the interdot distance is 30 nm, the dot Gaussian
confinement radius is 30 nm, and the effective central barrier
is 9.61 meV.  Again, the inset shows the magnetic field dependence 
of the splitting between the lowest singlet and triplet states
at three effective barrier heights of 3.38, 6.28, and 9.61 meV.}
}
\label{fig2}
\end{figure}

Our findings, shown in Fig.~\ref{fig2}(a), are rather striking: we
find that the six-electron Hilbert space (i.e. energy level spectra of
the two-dot system) is qualitatively different from
the two-electron double quantum dot case \cite{HD} 
shown in Fig.~\ref{fig2}(b) (included here for comparison),
and the multielectron system (with one electron
in the outermost open ``valence'' shell) does not necessarily behave
as a simple one effective spin per dot model.  We find that quantum
computation using quantum dot spin qubits and exchange gates will
most probably require the application of an external magnetic field 
or other means to
ensure a well defined sub-Hilbert space, 
which is an essential QC requirement \cite{DiVincenzo}, 
in the exchange-based two-qubit operations.

In the representative energy spectra presented in 
Fig.~\ref{fig2}(a) (with parameters in the figure caption), 
the S electrons are tightly confined to the individual QDs.
We include (for comparison) results of our restricted Hartree-Fock
ground state energies as dotted thick black lines. Notice that the CI
calculation produces a 3-4 meV improvement in the ground state
energy, mostly by introducing electron correlations to minimize
Coulomb repulsion.
The lowest lines at each field in Fig.~\ref{fig2}(a) actually 
consist of two
lines corresponding to the lowest singlet and triplet 
states. As their energy splitting is in the range of 0.01 to 
0.05 meV, the difference is too small to show up in this figure.  
Since this energy difference (singlet-triplet or exchange splitting)
is crucial in 
two-qubit operations, we plot the magnetic field 
dependence of the singlet-triplet splitting in 
the insets of Fig.~\ref{fig2}.  Notice that
the high magnetic field part of the inset
of Fig.~\ref{fig2}(a) is quite similar to 
that for the two-electron double dot case 
shown in the inset of Fig.~\ref{fig2}(b).
Here both triplet and singlet
states consist mainly of the lower energy P orbitals
$\psi_{LP-}$ and $\psi_{RP-}$;
the first subscript refers to the left or right QD, the
second refers to the orbital quantum number (S, P, or D)
of the Fock-Darwin state sequence,
and the third is the orbital magnetic quantum number.
Strong magnetic fields tightly squeeze the radii
of these P states so that their
overlap originates entirely
from their exponentially vanishing tails, 
leading to the similar high field behavior 
in Figs.~\ref{fig2}(a) and \ref{fig2}(b).
At low fields the exchange splitting in the multielectron system
has a much more complicated behavior than its 
single-electron counterpart.  At zero field the 
splitting is close to zero, implying a
delicate balance between electron kinetic energy and Coulomb
interaction.  The splitting quickly increases for lower 
central barriers as the outer-shell Hartree-Fock states change
quickly from an even superposition of $\psi_{P-}$ and $\psi_{P+}$
states to mainly $\psi_{P-}$ states.  For large barriers the
triplet state is the ground state even at zero field, in
analogy to, for example, the oxygen molecule. 

A crucial feature of our results, not obviously apparent from  
Fig.~\ref{fig2}, is the constituency of the lowest two 
states.  At zero magnetic field, the ground singlet is an equal 
superposition of the singlet states formed from the core and four 
different pairs of Fock-Darwin
states: $\psi_{PL-}$ and $\psi_{PR-}$,
$\psi_{PL-}$ and $\psi_{PR+}$, $\psi_{PL+}$ and $\psi_{PR-}$, and
$\psi_{PL+}$ and $\psi_{PR+}$. 
Therefore, if
initially in the single QD the outermost electron is in an
arbitrary superposition of orbital P states, then as the barrier 
between the two neighboring QDs is lowered, several 
low-lying excited states will inevitably get involved as we project
the initial state into a superposition of all the double-QD
eigenstates.  Indeed, using the four P states listed above one can
form four singlet and four triplet states, so that there
are seven energy parameters (neglecting the
splitting of any triplet state due to external fields).  Therefore,
in the most general case one has to manipulate seven different phases
to produce a swap---a formidable (if not completely intractable)
task.  As the magnitude of the external magnetic field increases, the
lowest singlet and triplet states become simpler, consisting mainly
the $\psi_{PL-}$  and $\psi_{PR-}$ states.  Thus if the initial single
QD outer shell electron state is purely $\psi_{P-}$, 
then only the lowest two states get involved 
as the interdot barrier is lowered, and the electron dynamics
is directly analogous to the original proposal of a single
electron confined in each QD.  In other words, the
orbital degrees of freedom for the outermost electrons are 
essentially frozen, so that the electron dynamics can be described
by a simple spin Hamiltonian---the Heisenberg Hamiltonian, and 
important two-qubit operations such as swap can be easily 
realized.

Another important feature of 
the spectra in Fig.~\ref{fig2}(a) is the splitting between the
lowest two states and the higher excited states (defining the
sharpness of the exchange sub-Hilbert space).  This splitting
is relatively small at zero and low fields, increases with the
field for a few Tesla, then gradually decreases again at higher
fields.  This means that there exists an optimal intermediate
magnetic field regime where the adiabatic condition necessary for
quantum computation can be most easily satisfied.  Indeed, this
optimal field regime is defined close to the $P+$ and $D-$ crossing
of the Fock-Darwin state sequence. 

Our results show that certain multielectron
cases, such as the situation of 3 electrons in each quantum dot 
in a 2-QD system, can be 
mapped on to the effective single-electron picture only at
intermediate external magnetic fields.  Essentially, the field
lifts the P state degeneracy so that a sufficiently large energy
gap opens up between the states involved in the exchange process
and higher excited states, and the 6-electron ground state is
formed from the single-dot 3-electron ground states.   
The energy gap is $\hbar 
\omega_C$ where $\omega_C$ is the cyclotron frequency and is linearly 
proportional to the magnetic field.  Thus, a 1 Tesla field will
lead to a 1.5 meV splitting, a large energy considering that the
exchange constant $J$ is typically of the order of 0.1 meV or 
smaller.  On the other hand,
at low (or zero) fields, there exists
a multitude of low-lying excited states due to the P state 
degeneracy, and the ground state
electronic wavefunctions are quite complicated.  At high fields, 
there are again relatively low energy excited states coming
from the lower energy D states.  Thus 
the adiabatic condition dictates that intermediate external fields
near the P-D crossing
provide the optimal operating condition for a multielectron 
quantum computer.

There are other means (not involving the application of an external
magnetic field) one can employ to break the degeneracy in the
P and higher excited states.  For example,
deformation of a circular quantum dot can lift the degeneracy 
in the P (and presumably all the higher excited) states,
thus facilitating a more reliable and accurate 
two-qubit operation.  If a circular
parabolic well is slightly deformed into an elliptical well, the
energy splitting between the two new P levels is $\frac{1}{2}\hbar
\omega_0 e$ where $e$ is the ellipticity.   Alternatively,
spin-orbit coupling can lift the orbital degeneracy, although it
would be quite small in our system, as we have 
only a few conduction electrons at the bottom of the GaAs conduction
band \cite{BLD,HD},  and it mixes the orbital and spin states, which
is what we try to avoid.

If we examine the physical picture underlying the effective qubit
behavior of the multielectron scenario closely, it is clear that a
crucial point is that the extra unpaired electron should not have
access to low-energy excited orbital states.  Thus, in general
multielectron coupled quantum dot systems do not reduce to a simple
Heisenberg exchange Hamiltonian in zero magnetic fields.  One might
speculate that a multielectron case may be analogous to the single
electron case when the number of electrons in a single dot is a full
shell minus one: 1, 5, 11, ..., $n(n+1)-1$, etc.  However,
particle-hole symmetry determines that in these cases pair-breaking
excitations in the outer shell will affect the low energy dynamics so
that these multielectron systems would actually be similar to the 3
electron (in a single dot) case we study here (and not to single
electron systems). We therefore conclude that multielectron circularly
symmetric quantum dot systems in zero external magnetic fields may
not be suitable as solid state spin qubits.  Thus one should either
use single electron quantum dots as in the original Loss-DiVincenzo
proposal (which may be a difficult task in practice) or apply
external magnetic fields (or break the circular symmetry using
controlled deformation) as we show in this paper.  The understanding
of multielectron systems as carried out in this paper may be an
important step in the realistic fabrication of spin-based QD-QC
architecture.

We conclude by emphasizing a general principle which is
explicitly demonstrated by the theoretical results presented in this
paper.  Quantum computation, in contrast to regular digital Boolean
classical computation, is analog, and the algorithm
is defined by the system Hamiltonian.  One must know the quantum
Hamiltonian (e.g. the exchange Hamiltonian in our QD-QC example)
controlling the qubit dynamics in the system accurately in order
to carry out meaningful quantum computation.  Our multielectron QD
calculations compellingly demonstrate the potential problems that may
arise---the effective single-electron Heisenberg exchange Hamiltonian
seems an eminently reasonable choice for QD-QC until
one looks carefully at the multielectron situation as we do here,
finding important qualitative differences with the effective
single-electron approach which can only be remedied through detailed
theoretical calculations.  We believe that the important lesson
presented in our example in this paper is quite generic: Know your
Hamiltonian well before you build your quantum computer. 

This work is supported by US-ONR and ARDA.

\end{document}